\documentclass[reprint,superscriptaddress,citeautoscript,floatfix,longbibliography,bibnotes,onecolumn]{revtex4-2}
\usepackage{times,siunitx,amsmath,amsfonts,amsthm,amssymb,amsbsy,braket,graphicx,hyperref,orcidlink,notes2bib}
\usepackage[utf8]{inputenc}
\usepackage[T1]{fontenc}
\usepackage[capitalize]{cleveref}
\hypersetup{pagebackref=true,hypertexnames=true,colorlinks=true,urlcolor=blue,linkcolor=blue,citecolor=red}
\DeclareMathOperator{\sgn}{sgn}
\DeclareMathOperator{\pf}{pf}
\newcommand{\ii}{\mathrm{i}}
\newcommand{\ee}{\mathrm{e}}

\begin{document}

\title{
Braiding with Majorana lattices: Groundstate degeneracy and supersymmetry
}
\author{Pasquale Marra\orcidlink{0000-0002-9545-3314}}
\email{pmarra@ms.u-tokyo.ac.jp}
\affiliation{
Graduate School of Mathematical Sciences,
The University of Tokyo, 3-8-1 Komaba, Meguro, Tokyo, 153-8914, Japan}
\affiliation{
Department of Physics, and Research and Education Center for Natural Sciences, 
Keio University, 4-1-1 Hiyoshi, Yokohama, Kanagawa, 223-8521, Japan}
\author{Daisuke Inotani\orcidlink{0000-0002-7300-1587}}
\author{Muneto Nitta\orcidlink{0000-0002-3851-9305}}
\affiliation{
Department of Physics, and Research and Education Center for Natural Sciences, 
Keio University, 4-1-1 Hiyoshi, Yokohama, Kanagawa, 223-8521, Japan}
\date{\today}

\begin{abstract}
Majorana-based topological qubits are expected to exploit the nonabelian braiding statistics of Majorana modes in topological superconductors to realize fault-tolerant topological quantum computation. Scalable qubit designs require several Majorana modes localized on quantum wires networks, with braiding operations relying on the presence of the groundstate degeneracy of the topologically nontrivial superconducting phase. However, this degeneracy is lifted due to the hybridization between Majorana modes localized at a finite distance. Here, we describe a braiding protocol in a trijunction where each branch consists of a lattice of Majorana modes overlapping at a finite distance. We find that the energy splitting between the groundstate and the lowest-energy state decreases exponentially with the number of Majorana modes if the system is in its topologically nontrivial regime. This result does not rely on the specific braiding geometry and on the details of the braiding scheme but is a consequence of the supersymmetry and nontrivial topology of the effective low-energy Hamiltonian describing the Majorana lattice.
\end{abstract}

\maketitle

\section{Introduction}

Topologically-protected Majorana zero modes localized at the boundaries of the nontrivial phase of a one-dimensional (1D) topological superconductor~\cite{kitaev_unpaired_2001,lutchyn_majorana_2010,oreg_helical_2010} are expected to exhibit nonabelian exchange statistics, which can be exploited to realize the quantum gates of a topological quantum computer~\cite{ivanov_non-abelian_2001,kitaev_fault-tolerant_2003,nayak_non-abelian_2008}.
Motivated by their potential impact in the field of quantum computing, there is an ongoing experimental effort to realize Majorana modes in realistic and scalable devices, including semiconducting nanowires with strong spin-orbit coupling proximitized by conventional superconductors, semiconductor-superconductor planar heterostructures, and in arrays of magnetic adatoms on a conventional superconductor substrate (see Refs.~\cite{frolov_topological_2020,flensberg_engineered_2021}).
In principle, the nonabelian braiding statistics can be probed by adiabatically exchanging Majorana modes in physical space in a trijunction~\cite{alicea_non-abelian_2011,clarke_majorana_2011,romito_manipulating_2012,halperin_adiabatic_2012}.
The exchange operations of a number $2N$ of Majorana modes are unitary transformations acting on the corresponding $2^N$-fold degenerate groundstate.
However, this groundstate degeneracy is exact only for Majorana modes localized at an infinite distance one from the other.
In the real world, however, Majorana modes gain a finite but exponentially small energy $\propto\ee^{-l/\xi_\mathrm{M}}$, where $l$ is the distance between the two Majorana modes and $\xi_\mathrm{M}$ their localization length.
This energy splitting is a source of quantum decoherence~\cite{knapp_dephasing_2018}, which is a fundamental hindrance to the realization of fault-tolerant topological quantum gates.
Moreover, scalable qubit designs require several Majorana modes localized on 2D or even 3D networks containing multiple trijunctions as building blocks~\cite{sau_universal_2010,alicea_non-abelian_2011,sau_controlling_2011,clarke_majorana_2011}.
Hence, it is crucial to determine the stability and robustness of the groundstate degeneracy in braiding designs containing several Majorana modes overlapping at a finite distance.

Recently, we proposed the realization of arrays of several partially-overlapping, zero-dimensional Majorana modes in proximitized nanowires via periodically-modulated magnetic fields superimposed by a rotating field in the same plane~\cite{marra_1d-majorana_2022,marra_dispersive_2022}.
This lattice of Majorana modes can realize a ``Majorana pump'' with the adiabatic translation of one Majorana mode for each half rotation of the externally applied field.
Remarkably, the groundstate degeneracy becomes exact in loop geometries or for an infinite number of Majorana modes~\cite{marra_1d-majorana_2022,marra_dispersive_2022}.
This exact degeneracy is a direct consequence of the quantum mechanical supersymmetry (SUSY) of the groundstate~\cite{hsieh_all-majorana_2016,huang_supersymmetry_2017,marra_1d-majorana_2022,marra_dispersive_2022}, which describes the symmetry between many-body states with different fermion parity, and which is generally present in topological superconductors~\cite{hsieh_all-majorana_2016,huang_supersymmetry_2017,grover_emergent_2014,rahmani_emergent_2015,rahmani_interacting_2019,rahmani_phase_2015} and their mechanical analogues~\cite{roychowdhury_supersymmetry_2022,lo_topology_2022}.
Here, by exploiting these ideas, we describe the realization of a braiding protocol in a trijunction, where each branch realizes a sliding lattice of Majorana modes overlapping at a finite distance, described by an effective $\mathbb{Z}_2$ topological invariant.
We find that, in the topologically nontrivial phase, the presence of several Majorana modes does not increase but exponentially suppresses the energy splitting, restoring the groundstate degeneracy in the limit of infinitely many Majorana modes.

\section{Trijunction with Majorana lattices}

\begin{figure}[t]
\includegraphics[width=\columnwidth]{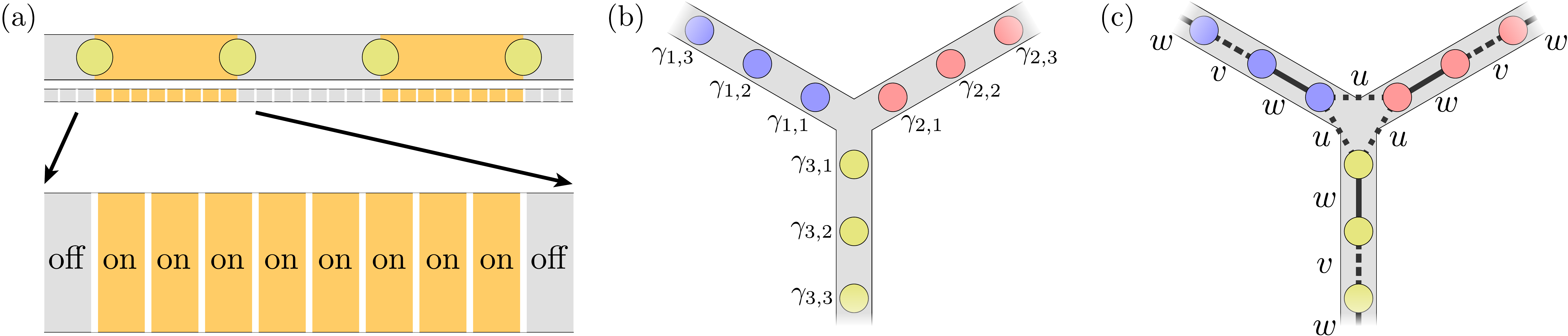}
\caption
{
(a)
Piano keyboard setup, where Majorana modes localized at the boundaries between trivial and nontrivial segments can be moved by manipulating magnetic textures or electric gates.
(b)
Majorana modes on a trijunction described by the effective Hamiltonian in \cref{eq:H1}.
(c)
Translational-invariant bipartite lattice with alternate couplings $w$ and $v$.
}
\label{fig:trijunction}
\end{figure}

We consider a T-shaped or Y-shaped trijunction with a lattice of $2N$ Majorana modes on each branch.
In order to obtain a lattice of regularly-spaced Majorana modes, one can employ periodical magnetic textures induced by nanomagnets~\cite{marra_1d-majorana_2022,marra_dispersive_2022}.
In this setup, the boundary between trivial and nontrivial segments can be smoothly controlled by rotating an applied field, inducing the adiabatic translation of the Majorana modes, i.e., and adiabatic ``Majorana pump''~\cite{marra_1d-majorana_2022,marra_dispersive_2022}.
Alternatively, regularly-spaced Majorana modes can be produced in the presence of arrays of tunable spin-valves~\cite{zhou_tunable_2019} or magnetic tunnel junctions~\cite{fatin_wireless_2016,matos-abiague_tunable_2017}, or periodic arrays of electric gates~\cite{alicea_non-abelian_2011,scheurer_nonadiabatic_2013,bauer_dynamics_2018}.
In these configurations, by switching on and off the spin-valves, magnetic tunnel junctions, or electric gates, one can drive different segments of the wire in and out of the topologically nontrivial phase to obtain a regular lattice of Majorana modes localized at the boundaries between trivial and nontrivial segments.
In this ``piano keyboard'' setup~\cite{alicea_non-abelian_2011,scheurer_nonadiabatic_2013,bauer_dynamics_2018}, the Majorana modes can slide by slowly rearranging the magnetic texture or the electric gates configurations, as shown in \cref{fig:trijunction}(a).
Alternatively, one can consider a Y-junction in planar Josephson junctions formed by depositing superconducting islands on top of a topological insulator, with Majorana modes localized at the superconducting vortex cores, which can be moved by applied currents, voltages, or phase differences~\cite{hegde_a-topological_2020}.

If the lengths of the trivial and nontrivial segments are comparable to the Majorana localization length, there is a small overlap between the Majorana modes wavefunctions.
In this case the low-energy Hamiltonian of the system can be obtained by projecting onto the subspace of Majorana operators, which in a trijunction configuration gives
\begin{equation}\label{eq:H1}
		\mathcal{H}_\text{eff}=\ii\Gamma H \Gamma^\intercal=
		\ii \left(	u_{1,2} \gamma_{1,1}\gamma_{2,1}	+	u_{2,3} \gamma_{2,1}\gamma_{3,1}	+	u_{3,1} \gamma_{3,1}\gamma_{1,1}	\right)
		+
		\ii \sum_{m=1}^3\sum_{n=1}^{2N-1}	w_{m,n} \gamma_{m,n}\gamma_{m,n+1}
		,
\end{equation}
where $\gamma_{m,n}$ are the Majorana operators on each branch $m=1,2,3$ and with $n=1,2,\ldots,2N$ counting outward from the center, as shown in \cref{fig:trijunction}(b), $\Gamma=(\gamma_{1,1},\gamma_{2,1},\gamma_{3,1},\ldots,\gamma_{1,2N},\gamma_{2,2N},\gamma_{3,2N})$ the vector of the Majorana operators, $u_{m,m'}\in\mathbb R$ the coupling between Majorana modes $\gamma_{m,1}$ and $\gamma_{m',1}$ at the center of the junction, and $w_{m,n}\in\mathbb R$ the couplings between contiguous Majorana modes $\gamma_{m,n}$ and $\gamma_{m,n+1}$ on each branch.
Let us also assume that all the couplings can be written as $w_{m,n}=E_0\ee^{-l_{m,n}/\xi_\mathrm{M}}$  in terms of the Majorana localization length $\xi_\mathrm{M}$ where $l_{m,n}$ is the distance between contiguous Majorana modes $\gamma_{m,n}$ and $\gamma_{m,n+1}$ on each branch, and $E_0$ a characteristic energy scale of the junction.
We will later consider a simpler case where each branch forms a translational-invariant bipartite lattice with $w_{m,n}=w$ or $w_{m,n}=v$ for $n$ odd and even, respectively, and with $u_{1,2}=u_{2,3}=u_{3,1}=u$, as in \cref{fig:trijunction}(c).

\section{Braiding protocol}

\begin{figure}[t]
\includegraphics[width=\columnwidth]{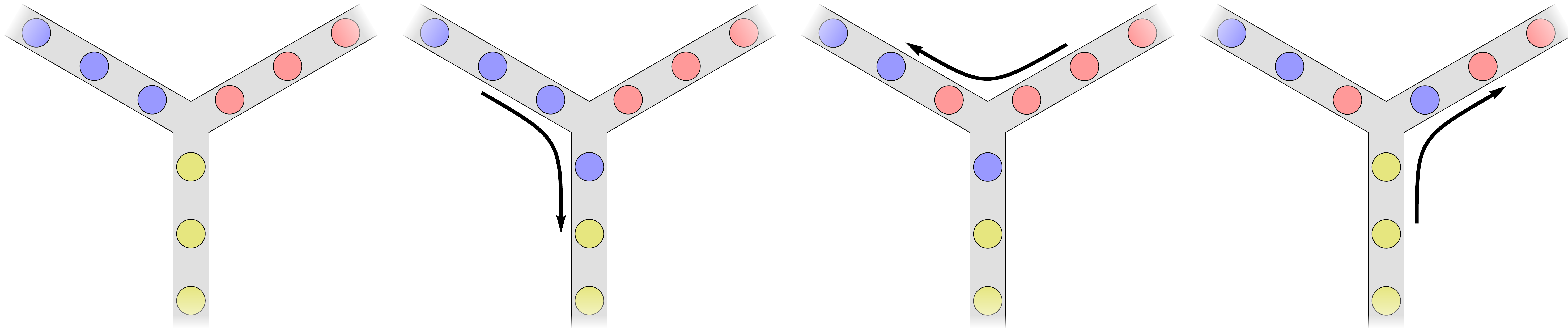}
\caption
{
Braiding of two Majorana modes in a trijunction obtained by sliding the Majorana modes at the boundaries between trivial and nontrivial segments. 
We move the Majorana modes on the upper left and lower branches downwards, then move the modes on the upper right and branches leftwards, and finally, move the modes on the upper right and lower branches upwards. 
}
\label{fig:braiding}
\end{figure}

The Majorana modes can move along the three branches of the trijunction by manipulating magnetic textures or electric gates, as already mentioned [see \cref{fig:trijunction}(a)].
A simple braiding protocol 
which exchanges two Majorana modes at the center of the junction
can be engineered by sliding the lattices of Majorana modes separately on the three different branches, as illustrated in \cref{fig:braiding}.
This braiding protocol is similar to the braiding protocol of Majorana modes in planar Josephson junctions introduced in Ref.~\onlinecite{hegde_a-topological_2020}.
As a first step, we slide the Majorana lattices on the upper left and lower branches in the downward direction, such that one Majorana mode from the upper left branch crosses the center of the junction, reaching the lower branch.
We then slide the Majorana lattices on the upper right and left branches in the leftward direction, such that one Majorana mode from the right branch crosses the center of the junction into the left branch.
Finally, we slide the Majorana lattices on the upper right and lower branches in the upward direction, such that one Majorana mode from the lower branch (which originally started in the upper left branch) crosses the junction into the right branch.
At the end of the day, the Majorana modes which were sitting close to the center of the junction on the upper left and right branches are exchanged.
Analogously, one can exchange the Majorana modes from the upper left and lower branches, and from the upper right and lower branches.
If performed adiabatically, these processes can be described as the unitary braiding operators 
$U_{m,m'}=\tfrac1{\sqrt2}\left(1+\gamma_{m,1}\gamma_{m',1}\right)=\exp{\left(\frac\pi4\gamma_{m1}\gamma_{m',1}\right)}$ with $m,m'\in\{1,2,3\}$,
which correspond to the adiabatic exchange of the two Majorana modes $\gamma_{m,1}$ and $\gamma_{m',1}$ at the center of the junction.

\section{Groundstate degeneracy and supersymmetry}

Intuitively, one may expect that the presence of many overlapping Majorana modes would spoil the topological protection by lifting the degeneracy of the groundstate. 
However, this is not the case.
To estimate the lowest energy level $E_1=\Delta E$ of the Hamiltonian above we notice that,
since the minimum of a list of non-negative numbers is less than or equal to the geometric average of the same list\bibnote{For a list of non-negative numbers $0\le x_1\le x_2\le\ldots\le x_n$, the geometric average is $\sqrt[n]{x_1x_2\cdots x_n}=\ee^{(\ln{x_1}+\ln{x_2}+\cdots+\ln{x_n})/n}\ge\ee^{(\ln{x_1}+\ln{x_1}+\cdots+\ln{x_1})/n}=x_1$.},
this energy level must be less than or equal to the geometric average of all the $3N$ positive energy levels (corresponding to $6N$ Majorana modes) of the Hamiltonian above, which yields $\Delta E\le(\prod_{n=1}^{3N} E_i)^{1/3N}$.
Let us then recall that the product of the all energy levels is equal to the determinant of the Hamiltonian matrix, which is an antisymmetric matrix, giving $|\det H|=|\pf H|^2$, and that, due to the particle-hole symmetry of the superconductor, every energy level has a particle-hole symmetric level with opposite energy, which gives $|\det H|=\prod_{n=1}^{3N} E_i^2$.
Hence, by explicitly calculating the the pfaffian of the Hamiltonian, one obtains
\begin{equation}\label{eq:geometric}
	\Delta E\leq\left(\prod_{i=1}^{3N} E_i\right)^{\frac1{3N}}=
	|\pf{H}|^{\frac1{3N}}=
	\left(	\prod_{m=1}^{3}\prod_{n=1}^{N}	w_{m,2n-1}	\right)^{\frac1{3N}}=
	E_0
	\ee^{	-\frac1{3N\xi_\mathrm{M}} \sum_{m=1}^{3}\sum_{n=1}^{N}	l_{m,2n-1}	}
	.
\end{equation}
This equation gives an upper bound to the energy splitting between the groundstate and the lowest-energy level. 
Surprisingly, the geometric average does not depend on the couplings between the three central Majorana modes $\gamma_{1,1}$, $\gamma_{2,1}$, and $\gamma_{3,1}$, and depends only on the couplings $w_{m,2n-1}$ between the contiguous Majorana modes $\gamma_{m,2n-1}$ and $\gamma_{m,2n}$, but not on the couplings $w_{m,2n}$ of the contiguous modes $\gamma_{m,2n}$ and $\gamma_{m,2n+1}$.
This mandates that, if the coupling between a single pair of contiguous Majorana modes $\gamma_{m,2n-1}$ and $\gamma_{m,2n}$ on one of the branches is zero, there exists at least one energy level which is exactly zero.
Hence, one can recover an exactly degenerate groundstate by tuning a single parameter.
The presence of a zero-energy level in a system of several Majorana modes coupled together is a manifestation of quantum mechanical SUSY~\cite{huang_supersymmetry_2017,marra_1d-majorana_2022,marra_dispersive_2022}.

\begin{figure}[t]
\includegraphics[width=\columnwidth]{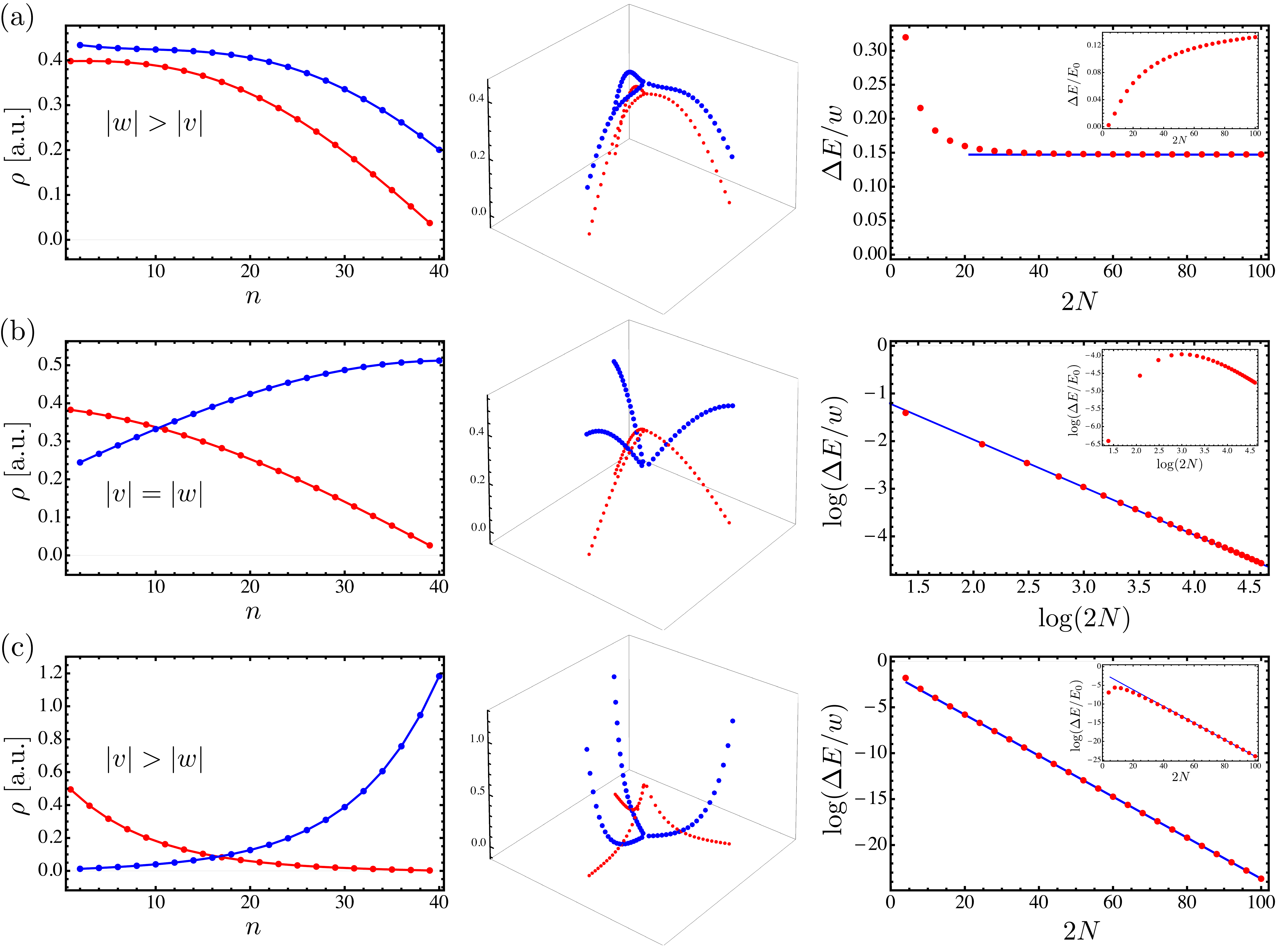}
\caption
{
Particle density $\rho$ of the lowest-energy state of a trijunction with branches of length $L=20\xi_\text{M}$ with $2N=40$ Majorana modes, plotted on a single branch and on the three branches, and corresponding energy splitting $\Delta E$ as a function of the number of Majorana modes $2N$. 
The data point corresponding to odd $\gamma_{m,2n-1}$ and even $\gamma_{m,2n}$ Majorana modes are grouped together.
(a)
Trivial phase $|w|>|v|$ ($w/v=1.2$).
The Majorana modes on odd and even sites hybridize along the entire width of the trijunction branches.
The energy splitting between the lowest-energy state and the groundstates increases and becomes proportional to the overlap $w$ between contiguous Majorana modes $\gamma_{m,2n-1}$ and $\gamma_{m,2n}$, increasing to a finite limit as $\Delta E\propto w\propto \ee^{-L/2N\xi_\mathrm{M}}$.
The continuous line at a constant value is a guide for the eye.
(b)
Topological phase transition $|v|=|w|$.
The Majorana modes on odd and even sites become partially decoupled forming two Majorana modes delocalized on the whole lattice.
In the limit $N\to\infty$, SUSY mandates that the $\Delta E\to0$, with each branch of the trijunction exhibiting two degenerate groundstates which are SUSY multiplets with opposite fermion parity.
As verified numerically,
the energy splitting decreases polynomially as $\Delta E\propto w/N$.
The continuous line is the best fit for $\Delta E/w\propto 1/N$.
(c)
Nontrivial phase $|v|>|w|$ ($w/v=0.8$).
The Majorana modes on odd and even sites decouple and hybridize into left and right Majorana end modes exponentially localized respectively at the center and at the outer ends of the three branches, with localization length $\xi_\text{eff}$.
The energy splitting decreases exponentially as $\Delta E\propto \ee^{-2N/\xi_\text{eff}}$.
The continuous line is the best fit for $\log(\Delta E/w)\propto -2N/\xi_\text{eff}$.
}
\label{fig:scaling}
\end{figure}

Let us consider the simpler case where each branch forms a translational-invariant bipartite lattice with $w_{m,n}=w=E_0\ee^{-l_w/\xi_\mathrm{M}}>0$ and $w_{m,n}=v=E_0\ee^{-l_v/\xi_\mathrm{M}}>0$ where $l_w$ and $l_v$ are the distances between contiguous Majorana modes $\gamma_{m,n}$ and $\gamma_{m,n+1}$ with $n$ odd and even, respectively, and with $u_{1,2}=u_{2,3}=u_{3,1}=u$, as in \cref{fig:trijunction}(c).
In this case the Hamiltonian in \cref{eq:H1} can be written as
\begin{equation}\label{eq:H2}
	\mathcal{H}_\text{eff}=\ii\Gamma H \Gamma^\intercal=
		\ii u\left(	\gamma_{1,1}\gamma_{2,1}	+	\gamma_{2,1}\gamma_{3,1}	+	\gamma_{3,1}\gamma_{1,1}	\right)
		+
		\ii\!\sum_{m=1}^3
		\left(
		\sum_{n=1}^{N}	w \gamma_{m,2n-1}\gamma_{m,2n}
		+\!
		\sum_{n=1}^{N-1}v \gamma_{m,2n}\gamma_{m,2n+1}
		\right)
		,
\end{equation}
The geometric average of all positive energy levels in this case becomes $|\pf{H}|^{\frac1{3N}}=w$, which gives an upper bound to the energy splitting $\Delta E\leq w$. 
This seems to suggest that, in a trijunction with branches of fixed length $L=Nl$ with $l=l_w+l_v$ and $\Delta l=l_w-l_v$, the energy splitting would increase to a finite limit as $\Delta E\propto w=E_0\ee^{-\Delta l/2\xi_\mathrm{M}}\ee^{-L/2N\xi_\mathrm{M}}$ with the number of Majorana modes $2N$.
As anticipated, this is not always the case, as we will show below.

The trijunction exhibits two topologically distinct phases realized respectively for $|w|>|v|$ and $|v|>|w|$, which correspond to the trivial and nontrivial phases of the three Majorana lattices of the three branches of the trijunction.
If the branches are decoupled $u=0$, indeed, each branch can be described as a bipartite lattice of 0D Majorana modes with an effective $\mathbb{Z}_2$ topological invariant $\mathcal{P}_\text{eff}=\sgn(|w|-|v|)$ (see Refs.~\cite{marra_1d-majorana_2022,marra_dispersive_2022}).
In the topologically trivial phase $|w|>|v|$, the energy splitting $\Delta E$ increases to a finite limit as $\Delta E\propto w\propto \ee^{-L/2N\xi_\mathrm{M}}$ as the number of Majorana modes increases $N\to\infty$ (as expected from the above argument).
At the topological phase transition $|v|=|w|$, the Majorana lattices exhibit two degenerate groundstates in the limit of infinite Majorana modes $N\to\infty$ (or, equivalently, in a finite lattice with periodic boundary conditions), which are SUSY multiplets with opposite fermion parity (the SUSY is spontaneously broken)~\cite{hsieh_all-majorana_2016,marra_1d-majorana_2022,marra_dispersive_2022}.
The two degenerate groundstates correspond to the vacuum and to counterpropagating and dispersive 1D Majorana modes delocalized on the whole length of the lattice~\cite{marra_1d-majorana_2022,marra_dispersive_2022}.
Hence, one expects that the energy splitting decreases with the number of Majorana modes, approaching zero for $N\to\infty$.
We verify numerically that the energy splitting decreases polynomially with the inverse of the number of Majorana modes as $\Delta E\propto w/N$ for $|v|=|w|$, even for finite $|u|>0$.
In the topologically nontrivial phase $|v|>|w|$ instead, each branch exhibits two Majorana end modes given by the hybridization of the 0D Majorana modes in each branch ${\widetilde\gamma}_\text{L}\propto\sum_j(w/v)^n \gamma_{m,2n-1}$ and ${\widetilde\gamma}_\text{R}\propto\sum_j(w/v)^{N+1-n} \gamma_{m,2n}$~\cite{marra_1d-majorana_2022,marra_dispersive_2022}. 
The left and right Majorana end modes, given by the superposition of 0D Majorana modes on odd and even sites, are localized at the opposite ends, i.e., respectively at the center and at the outer ends of the trijunction, with effective localization length $\xi_\text{eff}=l/|\log|w/v||=(l/|\Delta l|) \xi_\text{M}\ge \xi_\text{M}$.
If the coupling between the three branches is finite $|u|>0$, the three Majorana modes at the center hybridize into an unpaired Majorana mode at low energy~\cite{alicea_non-abelian_2011}.
The hybridization of the outer and inner Majorana modes yields a finite energy splitting $\Delta E\propto\ee^{-2N/\xi_\text{eff}}$ corresponding to the overlap of two end modes for each branch with localization length $\xi_\text{eff}$.
Remarkably, the presence of the inner and outer Majorana end modes and the exponential scaling of the energy splitting does not depend on the coupling $u$ between the three 0D Majorana modes at the center of the trijunction.
In particular, the numerical results indicate that $\Delta E\to0$ in the limit $N\to\infty$ for $|v|=|w|$, even for finite $|u|>0$.
This suggests that SUSY is not fully broken by the coupling of the three different branches of the trijunction.
However, it is an open question whether the trijunction topology further partially breaks the extended SUSY.

To numerically verify our findings, we diagonalize the Hamiltonian in \cref{eq:H2} considering a trijunction with fixed length $L$ and overlaps $w=E_0\ee^{-\Delta l/2\xi_\mathrm{M}}\ee^{-L/2N\xi_\mathrm{M}}$ and $v=E_0\ee^{+\Delta l/2\xi_\mathrm{M}}\ee^{-L/2N\xi_\mathrm{M}}$ with fixed $\Delta l$.
\Cref{fig:scaling} shows the particle density $\rho$ of the lowest-energy state and energy splitting $\Delta E$ as a function of the number of Majorana modes $2N$. 
In the trivial phase $|w|>|v|$ (i.e., $\Delta l>0$), Majorana modes on odd and even sites hybridize along the trijunction, and the energy splitting increases to a finite limit as $\Delta E\propto w\propto \ee^{-L/2N\xi_\mathrm{M}}$ [see \cref{fig:scaling}(a)].
At the topological phase transition $|v|=|w|$ (i.e., $\Delta l=0$), the Majorana modes on odd and even sites hybridize into two Majorana modes, and the energy splitting decreases polynomially as $\Delta E\propto w/N$ [see \cref{fig:scaling}(b)].
In the nontrivial phase $|v|>|w|$ (i.e., $\Delta l<0$), the Majorana modes on odd and even sites decouple and hybridize into left and right Majorana end modes, exponentially localized at the center and at the outer ends of the trijunction, with the energy splitting decreasing exponentially as $\Delta E\propto\ee^{-2N/\xi_\text{eff}}$ [see \cref{fig:scaling}(c)].
In all these regimes, we numerically verified that the asymptotic behavior of the energy splitting is independent on the choice of the coupling $u$ between the Majorana modes at the center of the junction.

The scaling of the energy splitting fully characterizes the different regimes of the trijunction:
Indeed, the energy splitting increases to a finite limit, decreases exponentially, or decreases polynomially, respectively, in the trivial, nontrivial, and at the topological phase transition.
One can also expect that this asymptotic behavior is preserved when the translational invariance is broken, i.e., in the presence of random disorder in the coupling between contiguous Majorana modes.

\section{Conclusions}

In this work, we analyzed the groundstate properties and described a braiding protocol for a trijunction where each branch contains an array of Majorana modes overlapping at a finite distance.
We found that, in this setup, the energy splitting between the groundstate and the lowest-energy many-body state decreases exponentially with the number of Majorana modes if the effective low-energy Hamiltonian of the Majorana lattices is topologically nontrivial.
This result does not depend on the geometry of the junction and on the details of the braiding scheme, and suggests that, in the nontrivial regime, the more Majorana modes there are in a topological qubit, the smaller the energy splitting.

\begin{acknowledgments}
P.~M.~is supported by the Japan Science and Technology Agency (JST) of the Ministry of Education, Culture, Sports, Science and Technology (MEXT), JST CREST Grant~No.~JPMJCR19T2 and Japan Society for the Promotion of Science (JSPS) Grant-in-Aid for Early-Career Scientists Grant~No.~20K14375.
M.~N.~is partially supported by the JSPS Grant-in-Aid for Scientific Research Grants~No.~JP18H01217 and No.~JP22H01221.
\end{acknowledgments}


\end{document}